\documentclass[article]{IEEEtran}
\usepackage{amsmath,amssymb,amsfonts,cuted}
\usepackage[final]{graphicx}
\usepackage{psfrag}
\usepackage{epsfig}
\usepackage[numbers,sort&compress]{natbib}
\usepackage{flushend}
\usepackage{array,booktabs}% http://ctan.org/pkg/{array,booktabs}

\usepackage{amsmath,amssymb,amsfonts}
\usepackage[final]{graphicx}
\usepackage{flushend}
\usepackage[numbers,sort&compress]{natbib}

\usepackage{ucs} 
\usepackage[utf8x]{inputenc}
\usepackage[usenames,dvipsnames]{xcolor}
\usepackage{tikz}
\usepackage{tkz-tab}
\usepackage{multirow}
\usepackage{latexsym}
\usepackage{amssymb}
\usepackage{amsmath}

\begin{document}

\title{Bivariate Rician shadowed fading model}
\author{J. Lopez-Fernandez, J.F. Paris~\IEEEmembership{Senior Member,~IEEE,} and~ E. Martos-Naya}% <-this % stops a space
\maketitle

\begin{abstract}
In this paper \footnote{This work has been submitted to the IEEE for possible publication. Copyright may be transferred without notice, after wich this version may no longer be accessible} we present a bivariate Rician shadowed fading model where the shadowing is assumed to follow a Nakagami-$m$ distribution. We derive exact expressions involving a single integral for both the joint probability density function (PDF) and the joint cumulative distribution function (CDF) and we also derive an exact closed-form expression for the moment generating function (MGF). As a direct consequence we obtain a closed-form expression for the power correlation coefficient between Rician shadowed variables as a function of the correlation coefficient between the underlying variables of the model. Additionally, in the particular case of integer $m$ we show that the PDF can be expressed in closed-form in terms of a sum of $m$ Meijer G-functions of two variables. Results are applied to analyze the outage probability (OT) of a dual-branch selection combiner (SC) in correlated Rician shadowed fading and the evaluation of the level crossing rate (LCR) and average fade duration (AFD) of a sampled Rician shadowed fading envelope.

\end{abstract}
\begin{IEEEkeywords}
Bivariate Rician shadowed, correlation, diversity reception, outage probability, level crossing rate, average fade duration.
\end{IEEEkeywords}

\section{Introduction}

The random fluctuation of the signal amplitude transmitted through a wireless channel has been extensively modeled in the literature \cite{simon2005}. Besides classical models like Rayleigh and Rice, more sophisticated statistical distributions are recalled to fit the behavior of the fading signal if accuracy is needed in more intricate scenarios. In \cite{Abdi2003} a modification of Rice model is presented where the amplitude of the Line-of-Sight (LOS) component is assumed to randomly fluctuate following a Nakagami-$m$ distribution. The resultant distribution is named Rician shadowed and fits to the land mobile satellite (LMS) channel experimental data \cite{Abdi2003} and has also proven to fit to the underwater acoustic channel (UAC) fading behavior \cite{Ruiz2012}. Two parameters characterize the Rician shadowed distribution. i.e. the Rician factor $K$ which denotes  the ratio between the average power in the LOS and the diffuse components and $m$ that describes the level of fluctuation of the LOS and ranges from 0.5 to $\infty$, were $m \rightarrow \infty$ means no fluctuation (constant LOS component). The proposed model includes as particular cases, the Rice ($m \rightarrow \infty$) and the Rayleigh ($K=0$) models.

In  \cite{Abdi2003}  closed-form expressions for the probability density function (PDF) and moment generating function (MGF) where presented while the cumulative density function (CDF) was derived in \cite{Paris2010}. This (and any other) univariate random model has applications on single input-single output systems but in the case of diversity combining systems the spatial distribution of reception antennas may cause the signals to present some degree of correlation which has motivated growing interest in the exploration of bi and multivariated distributions. Bivariated distributions have been profusely explored in \cite{Yacoub2007,Yacoub2008,Beaulieu2011,Yacoub2012, Lopez-Martinez2012NAKA,reig2014,Yacoub2016} including bivariate Rayleigh, Rician, Nakagami-m, Hoyt and $\kappa-\mu$ distributions among others. To the best of the author's knowledge the correlated Rician shadowed scenario has not yet been addressed in the literature. In this paper we present a bivariate Rician shadowed distribution model inspired by the approach presented in \cite{Tellambura2004} for correlated Rayleigh, Rician and Nakagami-m fading channels. 

In our work we obtain exact expressions for the PDF and CDF with one single integral involving the Kummer confluent hypergeometric function ${}_1F_1(\cdot,\cdot;\cdot)$ and derive a simple algebraic closed-form expression for the MGF. Moreover, for the case of integer $m$ we reach a closed-form expression for the PDF in terms of a finite sum of Meijer G-functions of two variables. The results are then used to analyze the outage probability (OT) of a dual-branch selection combiner (SC) in correlated Rician shadowed fading and to evaluate the level crossing rate (LCR) and average fade duration (AFD) of a sampled Rician shadowed fading envelope.

The remainder of this paper is organized as follows: in Section \ref{Bivariate Rice shadowed model} we introduce the statistical model which is then analyzed in Section \ref{Statistical analysis}. This section is subdivided in three subsections where the PDF, CDF and MGF are calculated separately. Application results are presented in Section \ref{Applications} whereas the main conclusions are exposed in Section \ref{Conclusions}.

\section{Bivariate Rician shadowed model}
\label{Bivariate Rice shadowed model}
The following notation will be used throughout the paper. We denote the absolute value and the expectation of a random variable $X$ as $|X|$ and $\mathbb{E}(X)$ respectively. The notation $X|Y$ stands for $X$ conditioned to $Y$. We write $X \sim \mathcal{N}_c(\mu,\sigma^2)$ to denote that random variable $X$ distributes as complex Gaussian with complex mean $\mu$ and variance $\sigma^2$. 

We start by modeling a set of correlated Rayleigh envelopes  
\begin{align}
G_k=&\sigma \sqrt{1-\rho} X_k + \sigma \sqrt{\rho} X_0, \;\;\;\; k=1,2,
\label{Eq:Gk}
\end{align}
where $X_k, k=1,2$ are independent random variables distributed as $\sim \mathcal{N}_c(0,\tfrac{1}{2})$ and $\rho$ is real number with  $0\leq \rho \leq 1$ \cite{Tellambura2004}. It can be easily checked that $\mathbb{E}(G_k)=0, $ $\mathbb{E}(G^2_k)=\sigma^2$ and that $\rho$ is the cross correlation coefficient between $G_1$ and $G_2$. Next we  model the fluctuation of the LOS component by incorporating a complex random variable $Z$ whose envelope is Nakagami distributed with real shaping factor $m \geq 0.5$ and $\mathbb{E}(|Z|^2)=\Omega_N$,  which results in our proposed expression for modeling correlated Rician shadowed random variables
\begin{equation}
H_k=G_k+Z=\sigma \sqrt{1-\rho} X_k + \sigma \sqrt{\rho} X_0+Z, \;\;\;\; k=1,2.
\label{Eq:Hk}
\end{equation}
In this model $G_k$ and $Z$ will account for the diffuse and LOS components respectively. Notice that the model assumes that the fluctuation of the LOS component (modeled by $Z$) is common to both variables $H_1$ and $H_2$ \footnote{In the practical application of this model (see section \ref{Applications}), $|H_1|$ and $|H_2|$ will account for the signal envelope level received either in one single antenna in two different time instants or in two different antennas at the same time. Assuming a common value of $Z$ means that the fluctuation of the LOS component varies in a larger time-space scale than that of the diffuse components.}.

The random variables $|H_k|$ are individually Rician shadowed distributed with $\mathbb{E}(|H_k|^2)=\sigma^2(1+K)$ for $k=1,2$ where $K=\Omega_N/\sigma^2$ is the Rician factor. The pair of random variables $|H_1|$ and $|H_2|$ defined in (\ref{Eq:Hk}) follow a bivariate Rician shadowed distribution, a fact that will symbolically be expressed as $\sim \bold{\mathcal{B}_{RS}}(\sigma,K,m,\rho)$.

%%%%%%%%%%%%%%%%%%%%%%%%%%%%%%%%%%%%%%%%%%%%%%%%% SECTION STATISTICAL ANALYSIS

\section{Statistical analysis}
\label{Statistical analysis}

In this section the PDF, CDF and MGF of the bivariate Rician shadowed model described in (\ref{Eq:Hk}) are derived. 

\subsection{Derivation of the PDF}

\subsubsection*{Lemma 1} Let $(R_1,R_2) \sim \bold{\mathcal{B}_{RS}}(\sigma,K,m,\rho)$ with $\sigma$, $K$, $m$, $\rho$ real positive, $m \geq 0.5$ and $0\leq \rho \leq 1$; then its joint PDF $f_{R_1,R_2}(r_1,r_2)$ is given by
\begin{multline}
f_{R_1,R_2}(r_1,r_2)= \tfrac{8\left( \frac{m \rho}{m \rho+K}\right)^m}{\sigma^6 \rho (1-\rho)^2} r_1 r_2\exp\left(-\frac{r_1^2+r_2^2}{\sigma^2(1-\rho)}\right) \\ \times \int_{0}^{\infty} \! x\exp\left(\tfrac{-(1+\rho)}{\sigma^2 \rho (1-\rho)}x^2\right) I_0\left( \frac{2r_1x}{\sigma^2(1-\rho)}\right)I_0\left( \frac{2r_2x}{\sigma^2(1-\rho)}\right) \\ \times {}_1F_1\left(m;1;\tfrac{K}{\sigma^2 \rho (\rho m +K)} x^2\right) dx,
\label{eq:int_final_PDF}
\end{multline}
where $I_0(\cdot)$ denotes the modified Bessel function of the first kind and zero order and ${}_1F_1(\cdot,\cdot;\cdot)$ is the Kummer confluent hypergeometric function \cite{gradshteyn2014}.
\begin{proof}
Let define $R_1=|H_1|$, $R_2=|H_2|$, $R_3=|V|$ and $R_4=|Z|$ where $V$ is defined as
\begin{equation}
 V=\sigma \lambda X_0+Z. 
\label{Eq:Q}
\end{equation}
The joint PDF of $R_1$ and $R_2$ can be expressed as a function of conditional PDFs as 
\begin{multline}
f_{R_1,R_2}(r_1,r_2)=\\
\int_{0}^{\infty} \!\!\int_{0}^{\infty} \!\! f_{(R_1,R_2)|R_3}(r_1,r_2,r_3)f_{R_3|R_4}(r_3,r_4)f_{R_4}(r_4)dr_3 dr_4.
\label{Eq:Procedimiento}
\end{multline}
The procedure consists in finding the successive conditional PDFs and then performing double integration. Using the definition of $V$ in (\ref{Eq:Q}) we can rewrite $H_k$ in (\ref{Eq:Hk}) as 
\begin{equation}
H_k=\sigma \sqrt{1-\rho} X_k + V, \;\;\;\; k=1,2.
\end{equation}
Consider now $V=v$ to be fixed. Then, the random variables $H_1$ and $H_2$ become independent (and so do $R_1$ and $R_2$) as they are a function of independent random variables $X_1$ and $X_2$ respectively. Moreover, both $R_1$ and $R_2$ turn into Rician variables with PDF \cite{simon2005}

\begin{equation}
f_{R_k|V}(r_k,v)=\frac{r_k}{\Omega^2} \exp\left(-\frac{r_k^2+|v|^2}{2\Omega^2}\right) I_0\left(\frac{|v|  r_k}{\Omega^2} \right),\;\;\;\; k=1,2,
\label{Rice_marginal_R1R2}
\end{equation}
where 
\begin{equation}
\Omega^2=\sigma^2(1-\rho)/2. 
\end{equation}
Notice that it is $|v|$ instead of the real or imaginary parts of $v$ what appears in the conditioned PDF in (\ref{Rice_marginal_R1R2}) so we can assess that $f_{R_k|V}(r_k,v)=f_{R_k|R_3}(r_k,r_3)$ for $k=1,2$. Next, since $R_1|V$ and $R_2|V$ are independent, their joint PDF can be written as the product of the marginal PDFs resulting in
\begin{multline}
f_{(R_1,R_2)|R_3}(r_1,r_2,r_3)=\frac{r_1 r_2}{\Omega^4} \exp\left(-\frac{r_1^2+r_2^2+2r_3^2}{2\Omega^2}\right) \\ \times I_0\left( \frac{r_1r_3}{\Omega^2}\right)I_0\left( \frac{r_2r_3}{\Omega^2}\right).
\label{Eq:PDF_Step1}
\end{multline}

Taking now $Z=z$ to be fixed, the random variable $R_3=|V|$ is also Rice distributed as can be seen from the definition in (\ref{Eq:Q}) and its PDF takes the form
\begin{equation}
f_{R_3|Z}(r_3,z)=\frac{r_3}{\Omega'^2} \exp\left(-\frac{r_3^2+|z|^2}{2\Omega'^2}\right) I_0\left(\frac{|z|r_3}{\Omega'^2}\right),
\label{Rice_marginal_R3}
\end{equation}
where
\begin{equation}
\Omega'^2=\sigma^2 \rho/2. 
\end{equation}
Since only $|z|$ takes part in (\ref{Rice_marginal_R3}) we can write $f_{R_3|Z}(r_3,z)=f_{R_3|R_4}(r_3,r_4)$ so that (\ref{Rice_marginal_R3}) can be expressed as 
\begin{equation}
f_{R_3|R_4}(r_3,r_4)=\frac{r_3}{\Omega'^2} \exp\left(-\frac{r_3^2+r_4^2}{2\Omega'^2}\right) I_0\left(\frac{r_3 r_4}{\Omega'^2}\right).
\label{Eq:PDF_Step2}
\end{equation}
Finally, the determination of $f_{R_4}(r_4)$ is straightforward as $R_4=|Z|$ is Nakagami distributed with parameters $m$ and $\Omega_N$, whose PDF is given by \cite{simon2005}

\begin{equation}
f_{R_4}(r_4)=\tfrac{2m^m}{\Gamma(m)\Omega_N^m}r_4^{2m-1} \exp\left( -\frac{m}{\Omega_N}r_4^2\right).
\label{Eq:PDF_Step3}
\end{equation}
Substituting (\ref{Eq:PDF_Step1}), (\ref{Eq:PDF_Step2}) and (\ref{Eq:PDF_Step3}) in (\ref{Eq:Procedimiento}) and reorganizing the double integral we get
\begin{multline}
f_{R_1,R_2}(r_1,r_2)=\tfrac{16\left(\frac{m}{K \sigma^2}\right)^m}{\Gamma(m) \sigma^6 \rho(1-\rho)^2}  r_1 r_2\exp\left(-\frac{r_1^2+r_2^2}{2\Omega^2}\right) \\ \times \int_{0}^{\infty} r_3\exp\left(- \tfrac{(1+\rho)}{\sigma^2 \rho (1-\rho)}r_3^2\right) I_0\left( \frac{r_1r_3}{\Omega^2}\right)I_0\left( \frac{r_2r_3}{\Omega^2}\right) I_{R_4}(r_3)dr_3, 
\label{eq:int_r3_r4_PDF}
\end{multline}
where $I_{R_4}(r_3)$ stands for the integral with respect to $r_4$ which is a function of $r_3$ 
\begin{equation}
 I_{R_4}(r_3)=\int_{0}^{\infty}  r_4^{2m-1} I_0\left( \frac{r_3r_4}{\Omega'^2}\right)  \exp\left( -\tfrac{m \rho +K}{\rho K \sigma^2 } r_4^2\right)  dr_4.
\label{Eq:I_{R_4}(r_3)}
\end{equation}
Using the identity $I_0(ax)= {}_0F_1\left(;1;\tfrac{a^2}{4}x^2\right)$ \cite{Wolfram0F1} where ${}_0F_1(;\cdot;\cdot)$ is the confluent hypergeometric function and using the integral \cite[eq. 5-7.522]{gradshteyn2014} a closed form solution can be obtained for $I_{R_4}(r_3)$, namely
\begin{equation}
 I_{R_4}(r_3)= \tfrac{\Gamma(m)}{2} \left(\tfrac{K\sigma^2  \rho}{\rho m +K}\right)^
m {}_1F_1\left(m;1;\tfrac{K}{\sigma^2 \rho (\rho m +K)} r_3^2\right).
\label{Eq:I_{R_4}(r_3)_closed}
\end{equation}
Substituting (\ref{Eq:I_{R_4}(r_3)_closed}) in (\ref{eq:int_r3_r4_PDF}) we get the proposed expression (\ref{eq:int_final_PDF}).\end{proof}
In case of integer $m$, expression (\ref{eq:int_final_PDF}) can be further manipulated so that a closed-form expression for the PDF can be obtained involving a sum of $m$ Meijer G-functions of two variables. 

\subsubsection*{Corollary 1}  Let $(R_1,R_2) \sim \bold{\mathcal{B}_{RS}}(\sigma,K,m,\rho)$ with $\sigma$, $K$, $\rho$ real positive, $0\leq \rho \leq 1$ and positive integer $m$; then its PDF can be expressed in closed form as
%
%
%
%\begin{multline}
%f_{R_1,R_2}(r_1,r_2)= \tfrac{4\pi^2}{\sigma^4(1-\rho^2)}r_1 r_2 \exp\left(-\frac{r_1^2+r_2^2}{\sigma^2(1-\rho)}\right) \times \\ G_{0,1:1,3:1,3}^{1,0:1,0:1,0}\left(\left. {- \atop 0} \right|\left. {1/2 \atop {0,0,1/2}} \right|\left. {1/2 \atop {0,0,1/2}} \right| \frac{\rho r_1^2}{2(1+\rho)}, \frac{\rho r_2^2}{2(1+\rho)} \right) ,\\ \;\;\;\;\; m =0
%\label{Eq:ClosedPDF1}
%\end{multline}

\begin{multline}
\label{Eq:ClosedPDF2}
f_{R_1,R_2}(r_1,r_2)=c_1 r_1 r_2 \exp\left(-\frac{r_1^2+r_2^2}{\sigma^2(1-\rho)}\right) \sum_{k=0}^{m-1}\frac{\binom{m-1}{k}c_2^k}{2 k!}  \\ \times G_{0,1:1,3:1,3}^{1,0:1,0:1,0}\left(\left. {- \atop {-k}} \right|\left. {1/2 \atop {0,0,1/2}} \right|\left. {1/2 \atop {0,0,1/2}} \right| c_3 r_1^2, c_3r_2^2 \right),
\end{multline}
where $G(\cdot)$ is the Meijer G-function of two variables (see II.13 in \cite{Nguyen1992}) and
\begin{align}
c_1=&\frac{8\pi^2 \left( \frac{m\rho}{m\rho+K}\right)^m}{\sigma^4 \rho(1-\rho)(2K-m(1-\rho))},\\
c_2=&\frac{K^2(1-\rho)}{\rho(\rho m+K)(2K-m(1-\rho))},\\
c_3=&\frac{K}{2(2K-m(1-\rho))}.
\end{align}

\begin{proof}
See Appendix A.
\end{proof}

%%%%%%%%%%%%%%%%%%%%%%%%%%%%%%%%%%%%%%%%%%%%% CDF %%%%%%%%%%%%%%%%%%%%%%%%%%%%%%%%

\subsection{Derivation of the CDF}
\subsubsection*{Lemma 2}  Let $(R_1,R_2) \sim \bold{\mathcal{B}_{RS}}(\sigma,K,m,\rho)$ with $\sigma$, $K$, $m$, $\rho$ real positive, $m \geq 0.5$ and $0\leq \rho \leq 1$; then its joint CDF $F_{R_1,R_2}(r_1,r_2)$ is given by

\begin{multline}
F_{R_1,R_2}(r_1,r_2)=\tfrac{2\left(\frac{m \rho}{m \rho +K} \right)^m}{\sigma^2 \rho}\int_{0}^{\infty}   x  \exp\left(- \frac{x^2}{\sigma^2 \rho} \right) \\ \times \left[ 1-Q_1\left( \frac{x}{\Omega} ,\frac{r_1}{\Omega} \right) \right] \left[ 1-Q_1\left( \frac{x}{\Omega}, \frac{r_2}{\Omega} \right) \right] \\ \times {}_1F_1\left(m;1;\tfrac{K}{\sigma^2 \rho(\rho m+K)} x^2\right) dx, 
\label{Eq:CDF}
\end{multline}
where $Q_{1}(\cdot,\cdot)$ is the first order Marcum $Q-$function.

\begin{proof}
We can proof Lemma 2 using the same approach employed for Lemma 1 replacing $f_{(R_1,R_2)|R_3}(r_1,r_2,r_3)$ by $F_{(R_1,R_2)|R_3}(r_1,r_2,r_3)$ in (\ref{Eq:Procedimiento}) which gives 

\begin{multline}
F_{R_1,R_2}(r_1,r_2)=\\
\int_{0}^{\infty} \!\!\int_{0}^{\infty} \!\!  F_{(R_1,R_2)|R_3}(r_1,r_2,r_3)f_{R_3|R_4}(r_3,r_4)f_{R_4}(r_4)dr_3 dr_4.
\label{Eq:Procedimiento_CDF}
\end{multline}

Both $R_1|R_3$ and $R_2|R_3$ are independent Rician random variables whose CDF is \cite{simon2005}

\begin{equation}
F_{R_k|R_3}(r_k,r_3)= \left[ 1-Q_1\left( \frac{r_3}{\Omega} ,\frac{r_k}{\Omega} \right) \right],\;\;\;\; k=1,2,
\label{Eq:Marginales_CDF}
\end{equation}
so we can write
\begin{multline}
F_{(R_1,R_2)|R_3}(r_1,r_2,r_3)= \\ \left[ 1-Q_1\left( \frac{r_3}{\Omega} ,\frac{r_1}{\Omega} \right) \right]\cdot \left[ 1-Q_1\left( \frac{r_3}{\Omega} ,\frac{r_2}{\Omega} \right) \right].
\label{Eq:Marginales_CDF}
\end{multline}

Substituting (\ref{Eq:PDF_Step2}), (\ref{Eq:PDF_Step3}) and (\ref{Eq:Marginales_CDF}) in (\ref{Eq:Procedimiento_CDF}) the integral with respect to $R_4$ is found to be identical to $I_{R_4}(r_3)$ in (\ref{Eq:I_{R_4}(r_3)}). Using the closed form result for $I_{R_4}(r_3)$ shown in (\ref{Eq:I_{R_4}(r_3)_closed}) we get the proposed CDF.
\end{proof}

%From the joint CDF in (\ref{Eq:CDF}) we can easily derive the marginal CDF of $R_1$ (or $R_2$) by taking the limit $F_{R_1}(r_1)=\lim_{r_2 \rightarrow \infty}F_{R_1,R_2}(r_1,r_2)$. Applying the fact that $\lim_{x \rightarrow \infty}Q_1(a,x)=0$, this yields
%
%\begin{multline}
%F_{R_k}(r_k)=\tfrac{2\left(\frac{m \rho}{m \rho +K} \right)^m}{\sigma^2 \rho}\int_{0}^{\infty}   x  \exp\left(- \frac{x^2}{\sigma^2 \rho} \right) \\ \times \left[ 1-Q_1\left( \frac{x}{\Omega}, \frac{r_k}{\Omega} \right) \right]  {}_1F_1\left(m;1;\tfrac{K}{\sigma^2 \rho(\rho m+K)} x^2\right) dx, \\ k=1,2.
%\label{Eq:CDF_marginal}
%\end{multline}

%%%%%%%%%%%%%%%%%%%%%%%%%%%%%%%%%%%%%%%%%%%%% MGF %%%%%%%%%%%%%%%%%%%%%%%%%%%%%%%%

\subsection{Derivation of the MGF}

\subsubsection*{Lemma 3}  Let $(R_1,R_2) \sim \bold{\mathcal{B}_{RS}}(\sigma,K,m,\rho)$ with $\sigma$, $K$, $m$, $\rho$ real positive, $m \geq 0.5$ and $0\leq \rho \leq 1$ and let $P_1=R_1^2$ and $P_2=R_2^2$; then the MGF of $P_1$ and $P_2$ is given by

\begin{multline}
\mathcal{M}_{P_1,P_2}(\theta_1,\theta_2)=\tfrac{K}{\sigma^2 \rho (m\rho+K)} \\ \times \frac{(a_1\theta_1 \theta_2+a_2\theta_1+a_3\theta_2+a_4)^{m-1}}{(b_1\theta_1 \theta_2+b_2\theta_1+b_3\theta_2+b_4)^m},
\label{Eq:MGF_lemma}
\end{multline}

where
\begin{align} 
a_1=&\tfrac{\sigma^2(1+\rho)(1-\rho)}{\rho},\;\;\; a_2=a_3=\tfrac{-1}{\rho},\;\;\;a_4=\tfrac{1}{\sigma^2\rho}, \\ b_1=&\tfrac{\sigma^2(1-\rho)}{\rho} \left(1+\rho-\tfrac{2K(1-\rho)}{m\rho+K}\right), \\ b_2=&b_3=-\frac{m+K}{\rho m+K},\;\;\;b_4=\frac{m\rho}{\sigma^2\rho(m\rho+K)}.
\end{align}

\begin{proof}
Following the same procedure used to proof Lemmas 1 and 2, we replace $f_{(R_1,R_2)|R_3}(r_1,r_2,r_3)$ by $\mathcal{M}_{(P_1,P_2)|R_3}(\theta_1,\theta_2,r_3)$ in (\ref{Eq:Procedimiento}) which gives

\begin{multline}
\mathcal{M}_{P_1,P_2}(\theta_1,\theta_2)=\\
\int_{0}^{\infty} \!\!\int_{0}^{\infty} \!\!  \mathcal{M}_{(P_1,P_2)|R_3}(\theta_1,\theta_2,r_3) f_{R_3|R_4}(r_3,r_4)f_{R_4}(r_4)dr_3 dr_4.
\label{Eq:Procedimiento_MGF}
\end{multline}

Since $P_1|R_3$ and $P_2|R_3$ are independent random variables their joint MGF can be written as the product of individual MGFs, this is  
\begin {equation}
\mathcal{M}_{(P_1,P_2)|R_3}(\theta_1,\theta_2,r_3)= \mathcal{M}_{P_1|R_3}(\theta_1,r_3)\cdot \mathcal{M}_{P_2|R_3}(\theta_2,r_3).
\label{Eq:MGF_conjunta_cond}
\end{equation}
Moreover, both $P_1|R_3$ and $P_2|R_3$ are non-central chi-square random variables whose MGF is readily obtained from  \cite[eq. 2-1-117]{proakis2001}. If we substitute this MGF in (\ref{Eq:MGF_conjunta_cond}) and then substitute (\ref{Eq:PDF_Step2}), (\ref{Eq:PDF_Step3})  and (\ref{Eq:MGF_conjunta_cond}) in the expression for the  MGF (\ref{Eq:Procedimiento_MGF}) and reorganize the double integral we can write

\begin{multline}
\mathcal{M}_{P_1,P_2}(\theta_1,\theta_2)=\tfrac{4\left(\frac{m}{\sigma^2 K}\right)^m}{\Gamma(m)\sigma^2 \rho} \prod_{k=1}^{2}\left[\frac{1}{1-\sigma^2 (1-\rho) \theta_k}\right] \\ \times \int_{0}^{\infty} \!\!\! r_3 \exp \left( -d_1 r_3^2 \right) I_{R_4}(r_3) dr_3,
\label{Eq:MGF_primera_integral}
\end{multline}
where
\begin{equation}
d_1=\frac{1}{\sigma^2 \rho}-\frac{\theta_1}{1-\sigma^2 (1-\rho) \theta_1}-\frac{\theta_2}{1-\sigma^2 (1-\rho) \theta_2},
\end{equation}
and where $I_{R_4}(r_3)$ is the same integral with respect to $R_4$ that appeared in the derivation of the PDF and CDF whose solution is shown in (\ref{Eq:I_{R_4}(r_3)_closed}). Substituting (\ref{Eq:I_{R_4}(r_3)_closed}) in (\ref{Eq:MGF_primera_integral}) we get

\begin{multline}
\mathcal{M}_{P_1,P_2}(\theta_1,\theta_2)=\tfrac{2}{\sigma^2 \rho} \left(\tfrac{\rho m}{\rho m +K}\right)^m \prod_{k=1}^{2}\left[\frac{1}{1-\sigma^2 (1-\rho) \theta_k}\right] \\ \times \int_{0}^{\infty} \!\!\! r_3 \exp \left( -d_1 r_3^2 \right) {}_1F_1\left(m;1;\tfrac{K}{\sigma^2 \rho (\rho m +K)} r_3^2\right) dr_3.
\label{Eq:MGF_segunda_integral}
\end{multline}

The integral with respect to $R_3$ can be solved in closed-form using again the integral \cite[eq. 5-7.522]{gradshteyn2014} yielding

\begin{multline}
\mathcal{M}_{P_1,P_2}(\theta_1,\theta_2)=\tfrac{1}{\sigma^2 \rho d_1} \left(\tfrac{\rho m}{\rho m +K}\right)^m \prod_{k=1}^{2}\left[\frac{1}{1-\sigma^2 (1-\rho) \theta_k}\right] \\ \times {}_2F_1\left(m,1;1; \tfrac{K}{\sigma^2 \rho (\rho m +K)d_1}\right). 
\label{Eq:MGF_prefinal}
\end{multline}
 
Finally, making use of the equivalence ${}_2F_1(a,1;1;z)=(1-z)^{-a}$ and after some algebraic simplification, (\ref{Eq:MGF_prefinal}) can be expressed in the final form presented in (\ref{Eq:MGF_lemma}).
\end{proof}

This result can be used to obtain a closed form expression that relates the correlation coefficient $\rho_{RS}$ of the square envelopes $P_1$ and $P_2$ of the bivariate Rician shadowed random variables to the correlation coefficient $\rho$ of the underlying model random variables. This is straightforward to verify since the correlation coefficient $\rho_{RS}$ is defined in terms of the moments of $P_1$ and $P_2$, namely

\begin{equation}
\rho_{BS}=\frac{\mathbb{E}[P_1P_2]-\mathbb{E}[P_1]\cdot\mathbb{E}[P_1]}{\sqrt{\mathbb{E}[P_1^2]-\mathbb{E}^2[P_1]}\cdot\sqrt{\mathbb{E}[P_2^2]-\mathbb{E}^2[P_2]}},
\label{Eq:rho_bs}
\end{equation}

and the moments in (\ref{Eq:rho_bs}) can obtained as derivatives of $\mathcal{M}_{P_1,P_2}(\theta_1,\theta_2)$ with the well known expressions

\begin{equation}
\mathbb{E}[P_k^n]=\left. \frac{\partial^n\mathcal{M}_{P_1,P_2}(\theta_1,\theta_2)}{\partial \theta_k^n}\right|_{\theta_k=0},\;\;\;\;k=1,2,
\label{Eq:DerivaMGF_1}
\end{equation}

\begin{equation}
\mathbb{E}[P_1P_2]=\left. \frac{\partial^2\mathcal{M}_{P_1,P_2}(\theta_1,\theta_2)}{\partial \theta_1 \partial \theta_2}\right|_{\theta_1=\theta_2=0}.
\label{Eq:DerivaMGF_2}
\end{equation}

\begin{figure}[t]
\centering
\includegraphics[width=.5\textwidth]{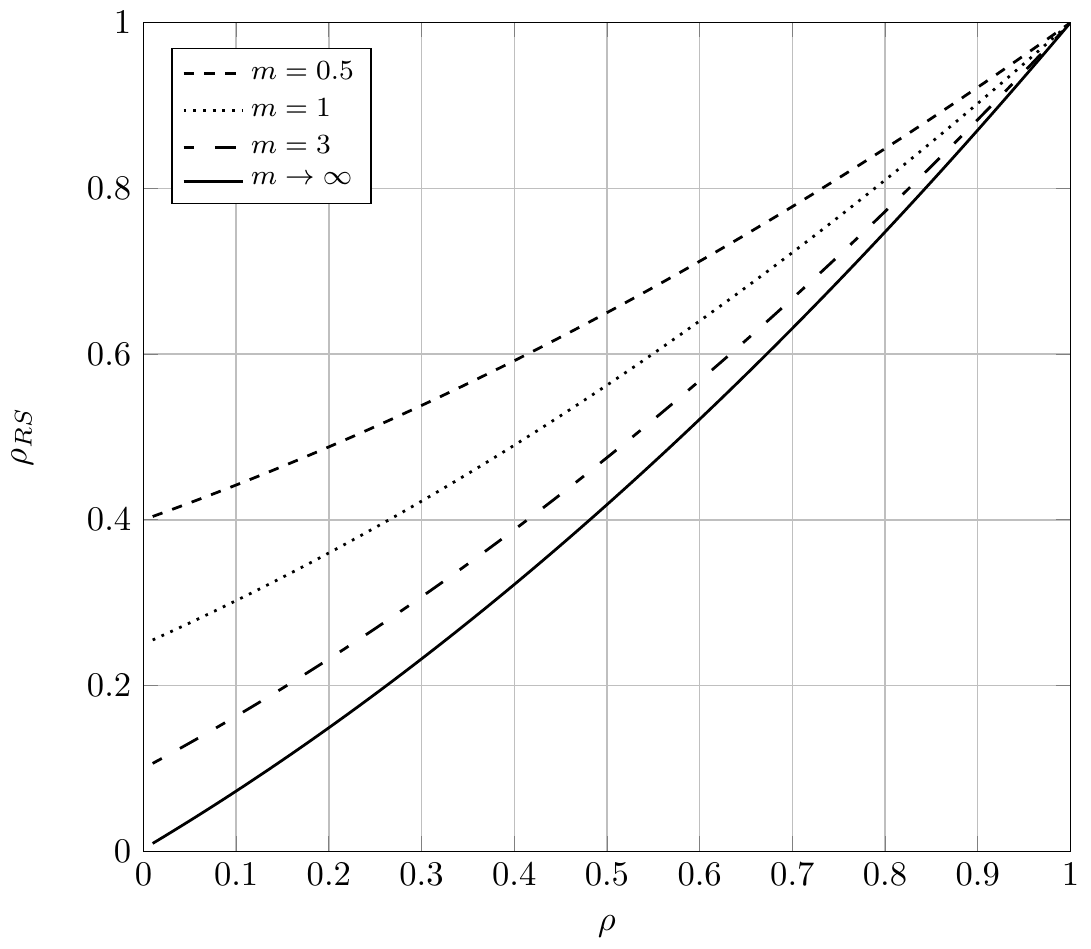}
\centering
\caption{Correlation coefficient $\rho_{RS}$ of square envelopes of the bivariate Rician shadowed random variables versus the correlation coefficient $\rho$ of the underlying model random variables for $K=1$ and different values of $m$.} 
\label{Fig_rho}
\end{figure}

Using (\ref{Eq:MGF_lemma}) in (\ref{Eq:DerivaMGF_1}) and (\ref{Eq:DerivaMGF_2}), and substituting the results in (\ref{Eq:rho_bs}) we reach a closed-form expression for $\rho_{BS}$ which is rather lengthy and is not shown here for simplicity. Fig. \ref{Fig_rho} depicts $\rho_{RS}$ versus $\rho$ for different values of $m$. According to the model in (\ref{Eq:Hk}), the correlation between the square envelopes $P_1$ and $P_2$ of $H_1$ and $H_2$ respectively has two origins. One is explicitly determined by the parameter $\rho$ and the other is a consequence of the common level of the LOS component fluctuation that we assumed to affect both variables. This explains the behavior of the plots in Fig. \ref{Fig_rho}. When $\rho \rightarrow 1$, $H_1 \rightarrow H_2$ which means $P_1 \rightarrow P_2$ and hence $\rho_{BS} \rightarrow 1$. However when $\rho \rightarrow 0$ there still remains a residual correlation between $P_1$ and $P_2$ due to the common level of fluctuation of the LOS component which makes $\rho_{BS}$ tend to a non-zero value. This value depends on the relative level of the LOS component fluctuation which is determined by both $m$ and $K$. i.e., see in Fig. \ref{Fig_rho} that $\rho_{BS}$ decreases as $m$ grows (less fluctuation of the LOS component) for any fixed $\rho$. For the limiting case of $m \rightarrow \infty$ (no fluctuation of the LOS component) see that $\rho_{BS} \rightarrow 0$ when $\rho \rightarrow 0$. Similarly, a reduction in the value of $K$ would also reduce the residual correlation for any fixed $m$. In this sense if we make $K \rightarrow 0$ all the curves on Fig. \ref{Fig_rho} will converge to the curve corresponding to $m \rightarrow \infty$.    

%%%%%%%%%%%%%%%%%%%%%%%%%%%%%%%%%%%%%%%%%%%%%%%%%%%%%%%%%%%%%%%%%%%%%%%%%    APPLICATIONS: OUTAGE %%%%%%%%%%%%%%%%%%%%%%%%%%%%%%%%%%%%%%%%%%%

\section{Applications}
\label{Applications}

Next we use the derived expression for the bivariate Rician shadowed CDF to analyze some interesting scenarios in communications starting with the outage probability (OT) study in dual-branch selection combining (SC) and following with second-order statistics of sampled Rician shadowed fading channels. 
\subsection{Outage probability of dual-branch SC}
In SC the receiver selects the branch with higher instantaneous SNR, $\gamma_k$, $k=1,2$ so the output SNR of the combiner is
\begin{equation}
\gamma_{SC}=\max(\gamma_1,\gamma_2)
\end{equation}
In the Rician shadowed fading scenario under study, the instantaneous SNR of the $k^{th}$  branch is
\begin{equation}
\gamma_k=\frac{R_k^2 E_s}{N_0}, \;\;\;\;\;k=1,2,
\end{equation}
while the average SNR is given by
\begin{equation}
\overline{\gamma}=\frac{\mathbb{E}[R_k^2] E_s}{N_0} =\frac{\sigma^2(1+K) E_s}{N_0}, \;\;\;\;\;k=1,2,
\label{Eq:AverageSNR}
\end{equation}
where $E_s$ is the symbol energy and $N_0$ is the noise spectral density which is assumed to be the same in both branches. Without loss of generality we will consider the normalization $E_s/N_0=1$ in the forthcoming expressions. 
The outage probability is defined as the probability that the instantaneous SNR $\gamma$ falls below a certain threshold $\gamma_{th}$. Using the previous definitions the outage probability in SC can be expressed as a function of the CDF in (\ref{Eq:CDF}) as
\begin{align}
P_{out}(\gamma_{th})=&\Pr(\gamma_{SC}<\gamma_{th})\\=&\Pr(\gamma_1<\gamma_{th},\gamma_2<\gamma_{th})\\=&\Pr(R_1^2<\gamma_{th},R_2^2<\gamma_{th}) \\ =&\Pr(R_1<\sqrt{\gamma_{th}},R_2<\sqrt{\gamma_{th}}) \\=&F_{R_1,R_2}(\sqrt{\gamma_{th}},\sqrt{\gamma_{th}}).
\end{align}

Using (\ref{Eq:CDF}) the expression for the outage probability is 

\begin{multline}
P_{out}(\gamma_{th})=\tfrac{2(1+K)\left(\frac{m \rho}{m \rho +K} \right)^m}{\overline{\gamma} \rho}\int_{0}^{\infty}   x  \exp\left(- \frac{(1+K)}{\overline{\gamma} \rho} x^2\right) \\ \times \left[ 1-Q_1\left( x \sqrt{\tfrac{2(1+K)}{\overline{\gamma}(1-\rho)}} ,\sqrt{\tfrac{2\gamma_{th}(1+K)}{\overline{\gamma}(1-\rho)}} \right) \right]^2 \\ \times {}_1F_1\left(m;1;\tfrac{K(1+K)}{\overline{\gamma}\rho(\rho m+K)} x^2\right) dx.
\label{Eq:outage}
\end{multline}

\begin{figure}[t]
\centering
\includegraphics[width=.5\textwidth]{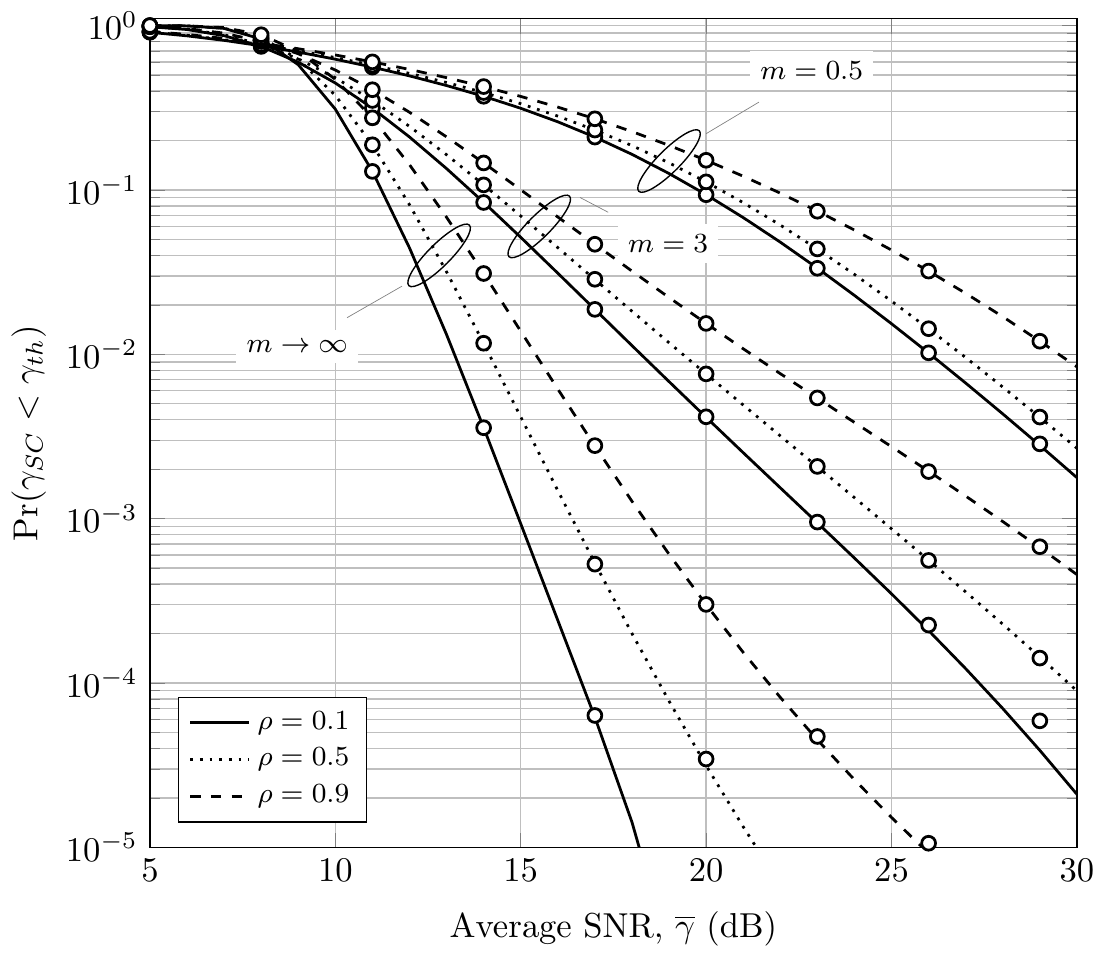}
\centering
\caption{Outage Probability as a function of the average SNR for $K=10$, and different values of $m$ and $\rho$. Lines correspond to theoretical  expressions and markers correspond to simulations} 
\label{Fig_out2}
\end{figure}

In Fig. \ref{Fig_out2} the outage probability with SC is depicted using numerical computation of (\ref{Eq:outage}) with MATLAB as a function of the average SNR $\overline {\gamma}$ for $\gamma_{th}=10\text{ dB}$, $K=10$, and different values of $m$ and $\rho$ . Notice that as $m$ grows, the  effect of the fluctuation of the LOS component diminishes yielding a decrease in outage probability for all the values of $\rho$ considered. The extreme case of no fluctuation of the LOS component ($m \rightarrow \infty$) corresponds to a correlated Rician fading scenario. The impact of different degrees of correlation between the signals arriving at the two branches can be also examined from Fig. \ref{Fig_out2}. Particularly, notice that the outage probability increases as the two branches correlate, e.g. as $\rho \rightarrow 1$ (which means $\rho_{BS} \rightarrow 1$) irrespective of the value of $m$. This is an expected behavior in SC scheme since the diversity gain decreases as the channels correlate (the limiting case $\rho=1$ leads to $\rho_{BS}=1$ which represents single-branch reception). On the opposite side, see that when $\rho$ decreases so does $\rho_{BS}$ (as stated in Fig. \ref{Fig_rho}) yielding a lower outage probability. Notice the total agreement between the theoretical and the simulation results in all instances.

%%%%%%%%%%%%%%%%%%%%%%%%%%%%%%%%%%%%%%%%%%%%%%%%%%%%%%%%%%%%%%%%%%%%%%%%%    APPLICATIONS: LCR %%%%%%%%%%%%%%%%%%%%%%%%%%%%%%%%%%%%%%%%%%%
\subsection{Level crossing rate and average fading duration}

Level crossing rate (LCR) and average fade duration (AFD) are second-order statistics that give information about the dynamics of the fading channels. The LCR is defined as the average rate at which the fading envelope crosses a certain threshold value while AFD measures the average time the envelope remains below a certain level. Traditionally they have been calculated following the approach proposed by Rice in \cite{Bell1945} which involves knowledge of the statistics of the continuous fading envelope and its time derivative. However in \cite{Lopez-Martinez2012} the authors proposed an interesting alternative formulation that takes into account the essential discrete-time nature of fading channels due to sampling. We will adopt this formulation in our study. In particular, let $R[n]$ be the envelope of a discrete random process obtained by sampling a continuous time random process with envelope $R(t)$ with sampling period $T_S$. The average rate at which $R[n]$ crosses a certain threshold level $u$ in the positive (or negative) direction will be denoted as $\text{LCR}(u)$ and can be analytically expressed as 

\begin{equation}
\text{LCR}(u)=\frac{\Pr\{R_1<u, R_2>u\}}{T_S},
\end{equation}
where $R_1 \triangleq R[n+1]$ and $R_2 \triangleq R[n]$. Noting that $\Pr\{R_1<u, R_2>u\}=\Pr\{R_1<u\}-\Pr\{R_1<u, R_2<u\}$ the LCR can be expressed as a function of the marginal CDF of $R_1$ and the joint CDF of $R_1$ and $R_2$ as
\begin{equation}
\text{LCR}(u)=\frac{F_{R_1}(u)-F_{R_1,R_2}(u,u)}{T_S}.
\label{Eq:LCR_def}
\end{equation}

The average time that the envelope $R_1[n]$ remains below a certain level $u$ will be denoted as $\text{AFD}(u)$ and can be calculated in terms of $\text{LCR}(u)$ as

\begin{equation}
\text{AFD}(u)=\frac{\Pr\{R_1<u\}}{\text{LCR}_R(u)}=T_S\left(\frac{F_{R_1}(u)}{F_{R_1}(u)-F_{R_1,R_2}(u,u)}\right).
\label{Eq:AFD_generico}
\end{equation}

In a Rician shadowed fading scenario, the marginal CDF $F_{R_1}(r_1)$ that appears in (\ref{Eq:LCR_def}) and (\ref{Eq:AFD_generico}) corresponds to the CDF of a single Rician shadowed random variable whose closed-form expression can be found in \cite[Eq. 8]{Paris2010}. Both \text{LCR}(u) and $\text{AFD}(u)$ can be hence computed using \cite[Eq. 8]{Paris2010} and (\ref{Eq:CDF}).

\begin{figure}[t]
\centering
\includegraphics[width=.5\textwidth]{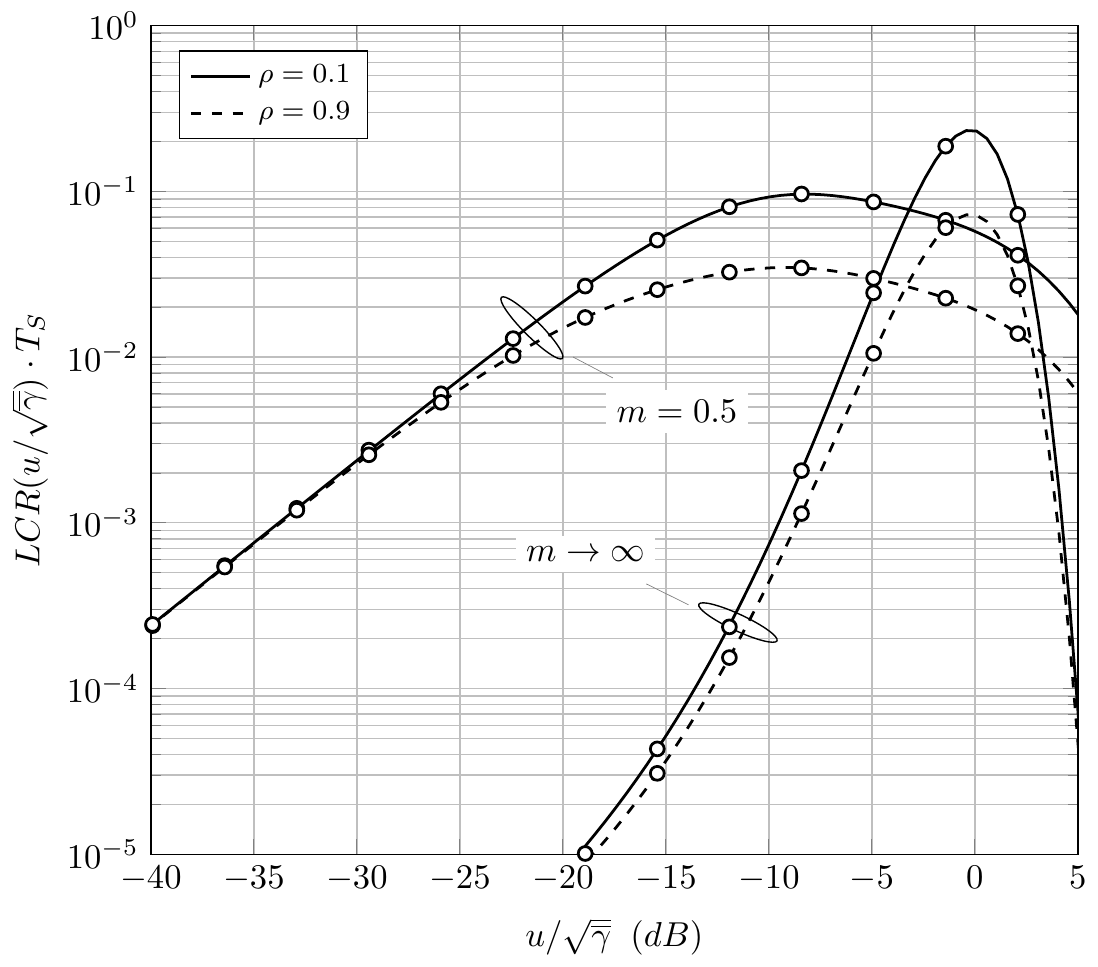}
\centering
\caption{LCR of the sampled Rician shadowed fading envelope versus normalized threshold level $u/\sqrt{\overline{\gamma}}$ for $K=10$ and different values of $m$ and $\rho$. Lines correspond to theoretical expressions and markers correspond to simulations} 
\label{Fig_lcr1}
\end{figure}

Fig. \ref{Fig_lcr1} shows the LCR normalized to $T_S$ as a function of the threshold level $u$ normalized to $\sqrt{\overline{\gamma}}$ for $K=10$ and different values of $\rho$ and $m$. The expected dependence of the LCR on the correlation between two consecutive samples is corroborated by the results shown in the figure. See that the LCR decreases as the correlation coefficient $\rho$ grows (whatever the value of $m$ chosen) since two consecutive envelope samples are more likely to take the same value. The extreme case of $\rho \rightarrow 1$ would yield equal sample values and therefore, no crossings, i.e. $\text{LCR} \rightarrow 0$. With regard to the effect of $m$ on the LCR notice that as the fluctuation of the LOS component decreases ($m$ grows) the LCR shape narrows around approximately the reference level (0 dB) and drops off quickly on both sides. This behavior is caused by the reduction in the dispersion of envelope values associated with a reduction in the level of fluctuation of the LOS component.

\begin{figure}[t]
\centering
\includegraphics[width=.5\textwidth]{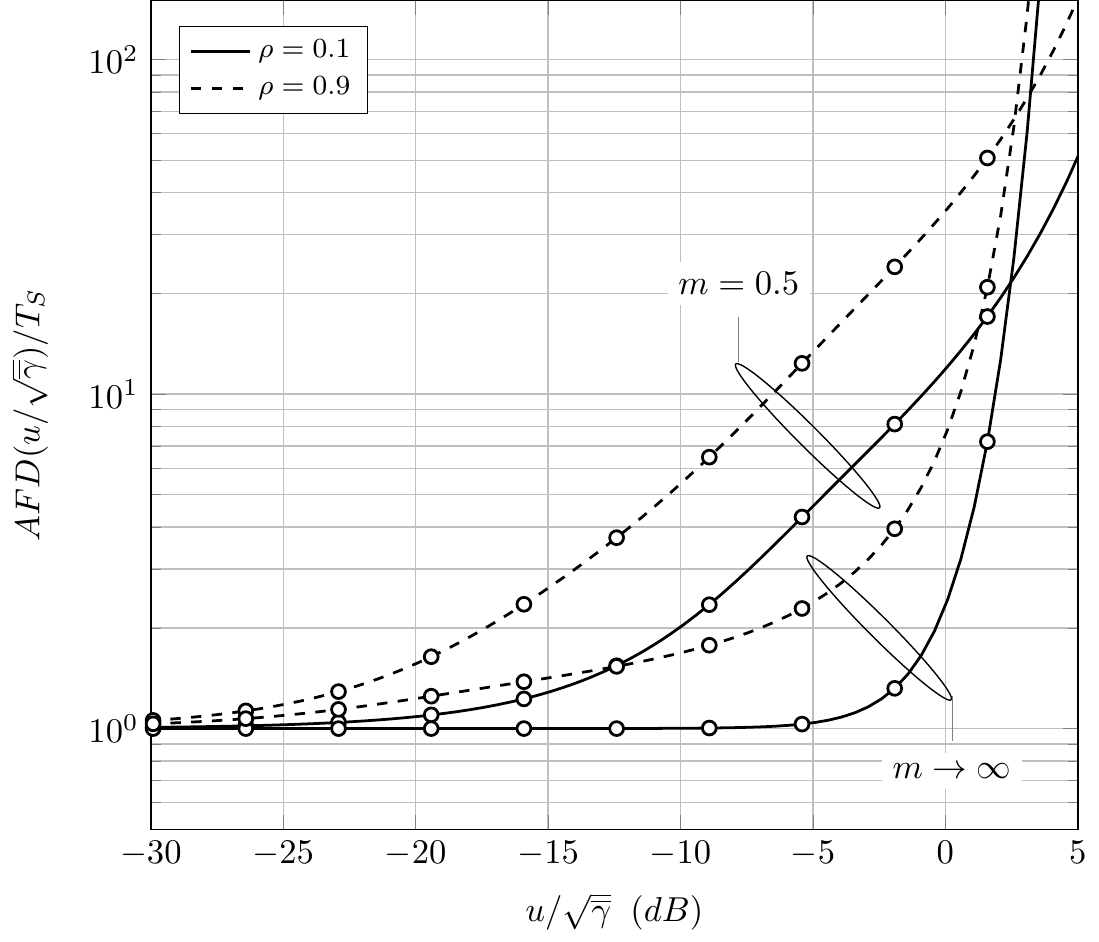}
\centering
\caption{AFD of the sampled Rician shadowed fading envelope versus normalized threshold level $u/\sqrt{\overline{\gamma}}$ for $K=10$ and different values of $m$ and $\rho$. Lines correspond to theoretical expressions and markers correspond to simulations} 
\label{Fig_afd1}
\end{figure}

In Fig. \ref{Fig_afd1} the AFD normalized to $T_S$ corresponding to the same parameters used in Fig. \ref{Fig_lcr1} is depicted as a function of the normalized threshold level $u/\sqrt{\overline{\gamma}}$. Opposite to what happens with the LCR, the AFD grows with $\rho$ and tends to infinity as $\rho \rightarrow 1$ for finite $u/\sqrt{\overline{\gamma}}$. See that all the curves of the normalized AFD tend to 1 (i.e. the de-normalized value of the AFD tends to $T_S$) as the threshold level decreases since the minimum expectable duration of a fading is one sample period. This lower bound in the value of the AFD is general for a sampled fading process. Again, theoretical and empirical results are in excellent agreement.

%%%%%%%%%%%%%%%%%%%%%%%%%%%%%%%%%%%%%%%%%%%%%%%%%% ECUACIONES LADO-LADO PUESTAS ANTES %%%%%%%%%%%%%%%%%%%%%%%%%%%%%%%%%%%%%%%%%%%%%%%%%%%%%%%%%%%

\begin{figure*}[!t]
%\setcounter{equation}{33}
% ensure that we have normalsize text
\normalsize
\begin{equation}
\label{Eq:Apendix_Triple_Meijer}
\begin{split}
\int_{0}^{\infty}& \! x\exp\left(\tfrac{-(1+\rho)}{\sigma^2 \rho (1-\rho)}x^2\right) I_0\left( \frac{2r_1x}{\sigma^2(1-\rho)}\right)I_0\left( \frac{2r_2x}{\sigma^2(1-\rho)}\right) {}_1F_1\left(m;1;\tfrac{K}{\sigma^2 \rho (\rho m +K)} x^2\right) dx = \\
 &\sum_{l=0}^{m-1} \frac{\pi^2 \binom{m-1}{l} \left( \tfrac{K}{\sigma^2 \rho (\rho m +K)} \right)^l}{2l!} \int_0^{\infty} x^l G_{0,1}^{1,0}\left(\left. {- \atop {0}} \right| \tfrac{2K-m(1-\rho)}{K\sigma^2(1-\rho)} x \right) \times \\
&G_{1,3}^{1,0}\left(\left. {1/2 \atop {0,0,1/2}} \right| \frac{r_1^2}{2\sigma^2(1-\rho)}x\right) G_{1,3}^{1,0}\left(\left. {1/2 \atop {0,0,1/2}} \right| \frac{r_2^2}{2\sigma^2(1-\rho)}x\right) dx.
\end{split}
\end{equation}
\hrule
\end{figure*}

%\begin{figure*}[!t]
%%\setcounter{equation}{34}
%% ensure that we have normalsize text
%\normalsize
%\begin{multline}
%\label{Eq:Apendix_Triple_Meijer_0}
%\int_{0}^{\infty} \! x\exp\left(\tfrac{-(1+\rho)}{\sigma^2 \rho (1-\rho)}x^2\right) I_0\left( \frac{2r_1x}{\sigma^2(1-\rho)}\right)I_0\left( \frac{2r_2x}{\sigma^2(1-\rho)}\right) {}_1F_1\left(0;1;\tfrac{K}{\sigma^2 \rho (\rho m +K)} x^2\right) dx = \\
%\tfrac{\pi^2}{2} \int_0^{\infty} G_{0,1}^{1,0}\left(\left. {- \atop {0}} \right| \tfrac{1+\rho}{\sigma^2 \rho (1-\rho)} x \right) G_{1,3}^{1,0}\left(\left. {1/2 \atop {0,0,1/2}} \right| \frac{r_1^2}{2\sigma^2(1-\rho)}x\right) G_{1,3}^{1,0}\left(\left. {1/2 \atop {0,0,1/2}} \right| \frac{r_2^2}{2\sigma^2(1-\rho)}x\right)  dx, \;\;\;\;\; m = 0
%\end{multline}
%\hrule
%\end{figure*}

\section{Conclusion}
\label{Conclusions}

In this paper a bivariate Rician shadowed model has been presented and exact expressions for the joint PDF, CDF and MGF have been derived. Closed-form expressions have been reached for the MGF and for the PDF, in this latter case for integer values $m$. As a derivation of these results a closed-form expression relating the power envelope correlation coefficient to the underlying correlation coefficient has been obtained. The CDF has been used to evaluate the outage probability for dual branch selection combining operating in correlated Rician shadowed fading channels and to analyze second-order statistics like the LCR and AFD for different levels of correlation and LOS component fluctuation. Simulation results agree with the proposed theoretical expressions.

%\appendix[Proof of the closed-form PDF expression in (\ref{Eq:ClosedPDF1}) and (\ref{Eq:ClosedPDF2})]
% or

%%%%%%%%%%%%%%%%%%%%%%%%%%%%%%%%%%%%%%%%%%% APPENDIX %%%%%%%%%%%%%%%%%%%%%%%%%%%%%%%%%%%%%%%%%%%%%%%%%%%%

%\appendices
\appendix[Proof of the closed-form PDF expression in (\ref{Eq:ClosedPDF2})]

\begin{proof}
We recall the expression that relates the function ${}_1F_{1}(m;1;z)$ to the Laguerre polinomials \cite{Wolfram1F1}, namely 
\begin{equation}
{}_1F_{1}(m;1;z)=\exp(z)L_{m-1}(-z), \;\;\;m\geq 1
\label{eq:1F1_Laguerre_m}
\end{equation}
where 
\begin{equation}
L_n(x)= \sum_{k=0}^{n}\binom{n}{k} \frac{(-1)^k}{k!}x^k
\label{eq:Laguerre}
\end{equation}
are the Laguerre polynomials. Then we take into account that the exponential function $\exp(x)$ and the modified Bessel function of the first kind and zero order $I_0(x)$ can be expressed \cite{Wolfram} as a particular case of the Miejer G-function of one variable defined in \cite[9.3]{gradshteyn2014}, namely
\begin{equation}
I_0(x)=\pi G_{1,3}^{1,0}\left(\left. {1/2 \atop {0,0,1/2}} \right| \frac{x^2}{4}\right)
\label{eq:I0_Meijer}
\end{equation}
and
\begin{equation}
\exp(x)= G_{0,1}^{1,0}\left(\left. {- \atop {0}} \right| -x\right)
\label{eq:exp_Meijer}
\end{equation}

If we first substitute (\ref{eq:1F1_Laguerre_m}) in (\ref{eq:int_final_PDF}) and then we make use of (\ref{eq:I0_Meijer}) and (\ref{eq:exp_Meijer}) the integral in (\ref{eq:int_final_PDF}) can be rewritten as a sum of $m$ integrals as shown in (\ref{Eq:Apendix_Triple_Meijer}) which involves the product of three Meijer G-functions. This integral has a closed-form solution \cite{Wolfram} which can be substituted in (\ref{eq:int_final_PDF}) and the final expression (\ref{Eq:ClosedPDF2}) is achieved after some basic yet cumbersome manipulation.
\end{proof}
\section*{Acknowledgment}
This work has been partially supported by FEDER and the Spanish and Andalusian
Governments, under Projects TEC2014-57901R, P11-TIC-8238 and P11-TIC-7109.

\ifCLASSOPTIONcaptionsoff
  \newpage
\fi



%\bibliographystyle{IEEEtran}
% argument is your BibTeX string definitions and bibliography databas\mathbb{E}(s)
%\begin{thebibliography}{1}
  %0.5em minus 0.4em\relax Harlow, England: Addison-Wesley, 1999.
%
%\end{thebibliography}

\bibliographystyle{ieeetr}
\bibliography{BivariateRicianShadowed_JLopezFdz}

%\begin{biography}[{\includegraphics[width=1in,height=1.25in,clip,keepaspectratio]{mshell}}]{Michael Shell}

%\end{IEEEbiography}

%\newpage


%\bibliographystyle{ieeetr}
%\makeatletter
%\renewcommand\@biblabel[1]{#1. }
%\makeatother


%\vfill


\end{document}